\documentclass[a4paper,11pt]{article}
\pdfoutput=1 
\usepackage{jheppub} 

\usepackage[utf8]{inputenc}

\usepackage[T1]{fontenc} 

\usepackage{Extrapackages}
\usepackage{xcolor}
\usepackage{cancel}
\usepackage{tensor}
\usepackage{simpler-wick}
\usepackage{tikz-feynman}
\tikzfeynmanset{compat=1.0.0}

\title{\boldmath{Mild Non-Gaussianities under Perturbative Control  \\ from Rapid-Turn Inflation Models}}

\author[1]{Theodor Bjorkmo,}
\author[2,3]{Ricardo Z.~Ferreira}
\author[2]{and M.C.~David Marsh}

\affiliation[1]{Department of Applied Mathematics and Theoretical Physics, University of Cambridge\\Wilberforce Road, Cambridge, UK}
\affiliation[2]{The Oskar Klein Centre, Department of Physics, Stockholm University, Stockholm 106 91, Sweden
}
\affiliation[3]{Nordita and KTH Royal Institute of Technology, \\Roslagstullsbacken 23, SE-106 91 Stockholm, Sweden}

\emailAdd{t.bjorkmo@damtp.cam.ac.uk}
\emailAdd{david.marsh@fysik.su.se}
\emailAdd{ricardo.zambujal@su.se}

\abstract{Inflation can be supported in very steep potentials if it is generated by rapidly turning fields, which can be natural in negatively curved field spaces. The curvature perturbation, $\zeta$, of these models undergoes an exponential, transient amplification around the time of horizon crossing, but can still be compatible with observations at the level of the power spectrum. However, a recent analysis (based on a proposed single-field effective theory with an imaginary speed of sound) found that the trispectrum and other higher-order, non-Gaussian correlators also undergo similar exponential enhancements. This arguably leads to 
`hyper-large' non-Gaussianities in stark conflict with observations, and even to the loss of perturbative control of the calculations. In this paper, we provide the first analytic solution of the growth of the perturbations in two-field rapid-turn models, and find it in good agreement with previous  numerical and single-field EFT estimates. We also show that the
nested structure of commutators of the in-in formalism has subtle and crucial consequences: accounting for these commutators, we show analytically that the naively leading-order piece (which indeed is exponentially large) cancels exactly in all relevant correlators. The remaining non-Gaussianities of these models are modest, and there is no problem with perturbative control from the exponential enhancement of $\zeta$. 
Thus, rapid-turn inflation with negatively curved field spaces remains a viable and interesting class of candidate theories of the early universe.

}

\begin{document}

  \maketitle
\flushbottom



%
%


\newcommand{\Mp}{M_{\mathrm{P}}}

\newcommand{\Nf}{{N_{\mathrm{f}}}}
\newcommand{\Lh}{\Lambda_{\mathrm{h}}}
\newcommand{\Lv}{\Lambda_{\mathrm{v}}}
\newcommand{\epsilonV}{\epsilon_{\mathrm{V}}}
\newcommand{\etaV}{\eta_{\mathrm{V}}}
\newcommand{\epsilonH}{\epsilon_{\mathrm{H}}}
\newcommand{\etaH}{\eta_{\mathrm{H}}}
\newcommand{\epsiloni}{\epsilon_{\mathrm{i}}}
\newcommand{\etai}{\eta_{\mathrm{i}}}

\newcommand{\db}[1]{\dot{\bar{#1}}}
\newcommand{\ddb}[1]{\ddot{\bar{#1}}}
\newcommand{\tsum}{{\textstyle{\sum}}}

\newcommand{\vv}[1]{\mathbf{#1}}
\newcommand{\vh}[1]{\mathbf{\hat{#1}}}
\newcommand{\nb}{\mathbf{\nabla}}
\newcommand{\dv}{\mathbf{\nabla}\cdot}
\newcommand{\cl}{\mathbf{\nabla}\times}
\newcommand{\bs}[1]{\boldsymbol{#1}}
\newcommand{\vvv}{\mathbf{v}}
\newcommand{\vvu}{\mathbf{u}}
\newcommand{\vvx}{\mathbf{x}}
\newcommand{\vvk}{\mathbf{k}} 
\newcommand{\vvp}{\mathbf{p}} 
\newcommand{\vvr}{\mathbf{r}}
\newcommand{\vvR}{\mathbf{R}} 
\newcommand{\vvG}{\mathbf{G}} 
\newcommand{\mc}[1]{\mathcal{#1}} 
\newcommand{\vvrd}{\dot{\mathbf{r}}}
\newcommand{\vvrdd}{\ddot{\mathbf{r}}}
\newcommand{\rrd}{\dot{r}}
\newcommand{\rrdd}{\ddot{r}}
\newcommand{\thetad}{\dot{\theta}}
\newcommand{\thetadd}{\ddot{\theta}}
\newcommand{\phid}{\dot{\phi}}
\newcommand{\phidd}{\ddot{\phi}}

\newcommand{\eq}[1]{\begin{equation}{#1}\end{equation}}
\newcommand\nt{\addtocounter{equation}{1}\tag{\theequation}}
\newcommand{\mui}{^{\mu}}
\newcommand{\mli}{_{\mu}}
\newcommand{\nui}{^{\nu}}
\newcommand{\nli}{_{\nu}}
\newcommand{\pd}[2]{\frac{\partial {#1}}{\partial {#2}}}
\newcommand{\td}[2]{\frac{d{#1}}{d{#2}}}
\newcommand{\spd}[2]{\frac{\partial^2 {#1}}{\partial {#2}^2}}
\newcommand{\std}[2]{\frac{d^2 {#1}}{d{#2}^2}}

\newcommand{\tb}[1]{\textcolor{purple}{#1}}

\definecolor{dmgreen}{rgb}{0.90,0.50,0.10}
\newcommand{\dmc}[1]{\textcolor{dmgreen}{DM:~#1}}
\newcommand{\rzf}[1]{\textcolor{green}{RZF:~#1}}
\newcommand{\be}{\begin{equation}}
\newcommand{\ee}{\end{equation}}
\newcommand{\beq}{\begin{equation}}
\newcommand{\eeq}{\end{equation}}

\newcommand{\Hint}{{\hat{H}_\text{int}}}

\newcommand{\Mnn}{\mu_n}
\newcommand{\Mns}{\mu_\times}
\newcommand{\Mss}{\mu_s}
\newcommand{\Vv}{V_{v}}
\newcommand{\Vvv}{V_{vv}}
\newcommand{\Vvw}{V_{vw}}
\newcommand{\Vww}{V_{ww}}

\section{Introduction}

Over the last few years there has been much interest in two-field inflation models with strongly non-geodesic motion \cite{Brown:2017osf,Mizuno:2017idt,Bjorkmo:2019aev,Bjorkmo:2019fls,Christodoulidis:2018qdw,Cremonini:2010ua,Renaux-Petel:2015mga,Renaux-Petel:2017dia,Garcia-Saenz:2018ifx,Garcia-Saenz:2018vqf,Grocholski:2019mot, Fumagalli:2019noh,Christodoulidis:2019jsx,Christodoulidis:2019mkj,Bravo:2019xdo,Easson:2007dh,Achucarro:2010jv,Achucarro:2010da, Achucarro:2012sm,Achucarro:2012yr,Cespedes:2012hu,Hetz:2016ics,Chen:2018uul,Chen:2018brw, Aragam:2019khr,Garcia-Saenz:2019njm, Achucarro:2019pux, Achucarro:2019lgo, Chakraborty:2019dfh}.
These models have been studied under various names (such as spinflation, hyperinflation, side-tracked inflation, angular inflation, and effective single-field theories with a reduced speed of sound from heavy fields),
but have recently been shown to belong to a general class of solutions known as `rapid-turn attractors' \cite{Bjorkmo:2019fls}. Intriguingly, inflation models with rapidly turning fields can be realised in potentials that are much too steep for standard slow-roll inflation \cite{Hetz:2016ics,Achucarro:2018vey},
and may thereby ameliorate the so-called inflationary `$\eta$-problem' \cite{Lyth:1998xn, Baumann:2014nda}. Moreover, under certain conditions, standard slow-roll inflation in hyperbolic field spaces can become unstable, triggering a `geometric destabilisation' \cite{Renaux-Petel:2015mga, Brown:2017osf} to the rapid-turn attractor, which is also known to arise as global attractor solution in hyperbolic field spaces
 \cite{Brown:2017osf,Mizuno:2017idt,Bjorkmo:2019aev,Christodoulidis:2018qdw,Renaux-Petel:2015mga,Renaux-Petel:2017dia,Garcia-Saenz:2018ifx,Cicoli:2018ccr, Grocholski:2019mot}. Given the prevalence of negative field space geometries in string compactifications (cf.~e.g.~\cite{Ooguri:2006in, Cicoli:2019ulk}), and the possibility of tackling the $\eta$-problem, these models are very attractive, and it is important to understand their observational viability.

Not only do these models have interesting background solutions, but their primordial perturbations are  also  unusual and intriguing. If the entropic mass is sufficiently large (which is not generally the case), these theories can be described by an  effective field theory (EFT) of 
a single field with a reduced speed of sound, $c_s$ (see e.g. \cite{Achucarro:2010jv,Achucarro:2010da,Achucarro:2012sm,Achucarro:2012yr,Cespedes:2012hu}). However, if the entropic mass is below a certain critical value, the perturbations undergo a transient instability before horizon crossing and have been proposed to be described by a single-field EFT with an \textit{imaginary} speed of sound \cite{
,Garcia-Saenz:2018ifx,Garcia-Saenz:2018vqf,Fumagalli:2019noh,Garcia-Saenz:2019njm}. This instability, first noticed in \cite{Cremonini:2010ua} and further developed in the context of hyperinflation in \cite{Brown:2017osf,Mizuno:2017idt}, causes the perturbations to grow exponentially before horizon crossing, leading to a suppressed tensor-scalar ratio and enhanced non-Gaussianity in the flattened configuration, with numerical results confirming the EFT predictions \cite{Fumagalli:2019noh}.

Beyond the bispectrum, however, this exponential growth has been argued to be disastrous. Reference \cite{Fumagalli:2019noh} recently 
 used the single-field EFT  to show that so-called scalar exchange diagrams appear to provide enormous contributions to the trispectrum and higher-order, non-Gaussian correlators. These theories would then predict exponentially large non-Gaussianities (in $g_{\rm NL},~\tau_{\rm NL}$ etc), leading to
  a strong tension with observations and even a loss of perturbative computational control. If true, this would rule out controlled rapid-turn models of inflation with strong non-geodesic motion, such as hyperinflation with a large turn-rate. 
 
 Fortunately however, as we will show, the situation is not so bleak.  In fact, we will demonstrate that when carefully accounting for the subtle structure of commutators in the in-in formalism, the apparently large and dominating contributions cancel out exactly. Estimates that do not account for the structure of nested commutators in the contributions to the correlation functions will therefore exponentially overestimate the resulting non-Gaussianity. This subtle cancellation effect is already known in the context of axially coupled gauge-fields during inflation {\cite{Ferreira:2015omg}, where the gauge field mode functions behave similarly. 
Upon adjusting the estimates of \cite{Fumagalli:2019noh} to account for the commutators, we find that these models of rapid-turn inflation are well within the perturbative regime, and do not lead to exponentially large non-Gaussianities. 
 

We close this paper with a general, analytic WKB computation of the growth of the perturbations in the two-field theory with strong turning. Our result agrees with the numerical analysis of  \cite{Mizuno:2017idt}, and gives further evidence for the correctness of the single-field effective field theory with imaginary speed of sound developed and used in \cite{Garcia-Saenz:2018ifx,Garcia-Saenz:2018vqf,Fumagalli:2019noh}.



We conclude that inflationary models involving rapidly turning fields remain observationally viable and, given their other theoretical strengths, provide appealing candidate models for the early universe.

\section{Rapid-turn inflation and the fear of hyper-large non-Gaussianities}

In this section, we briefly review the background dynamics of rapidly turning two-field solutions,  and illustrate this class of models by using hyperinflation as a particular example. We furthermore discuss the enhancement of the curvature around horizon crossing, and review why one may fear that the tri-spectrum and non-Gaussianity in higher-order correlators become exponentially enhanced in these solutions, following reference  \cite{Fumagalli:2019noh}.

\subsection{The rapid-turn attractor}
Two-dimensional field spaces admit, in addition to the standard  slow-roll inflation, a novel and interesting class of rapidly turning inflationary solutions. 
%
The background evolution of the fields is governed by the Klein-Gordon equation,
\eq{
\mathcal D_t\dot\phi^a+3H\dot\phi^a+G^{ab}V_{;b}=0,
}
where $\mathcal D_t X^a\equiv \dot X^a+\Gamma^a_{bc}\dot\phi^bX^c$, and $\Gamma^a_{bc}$ are the Christoffel symbols of the field space metric $G_{ab}$. 
We can drastically simplify this system by projecting the equations of motion onto the vielbein basis
\eq{
e^I_a=(v_a,~w_a),
}
where $v_a=V_{;a}/\|V_{;a}\|$ and $w_a$ is a (co-)vector field orthonormal to $v_a$, such that we work with the velocities $\dot\phi_v=v_a\dot\phi^a$ and $\dot\phi_w=w_a\dot\phi^a$. In this basis, the equations of motion for a homogeneous background of the fields become
 \cite{Bjorkmo:2019fls}
\eq{
\ddot\phi_v=-3H\dot\phi_v-\Vv+\Omega_v\dot\phi_v,\hspace{1cm}\ddot\phi_w=-3H\dot\phi_w-\Omega_v\dot\phi_v
\label{eq:bkgEoM}
}
where $\Vv=\|V_{;a}\|=v^aV_{;a}$, $\Vvw=v^aw^aV_{;ab}$ etc, and $\Omega_v=w_a\mathcal D_t v^a=(\Vvw\dot\phi_v+\Vww\dot\phi_w)/\Vv$ is the turn rate of the basis vectors. This turn-rate is in general different from the (dimensionless) turn-rate of the fields themselves, given by
\eq{
\omega=\|\mathcal D_t(\dot\phi^a/\dot\phi)\|/H \, ,
}
where $\dot \phi = \| \dot \phi^a \|$.

Equation \eqref{eq:bkgEoM} admits inflationary solutions. The standard slow-roll solution is characterised by 
small inflationary slow-roll parameters ($\epsilon,|\eta|\ll1$), and small accelerations of the fields  ($\ddot\phi_I=\mathcal O(\epsilon)H\dot\phi_I$), which can be achieved in sufficiently flat potentials with no rotation ($\omega=0$). However, even steep potentials can support inflation (still with $\epsilon,|\eta|\ll1$ and $\ddot\phi_I=\mathcal O(\epsilon)H\dot\phi_I$)
if  the turn rate is large compared to the slow-roll parameter, $\omega^2\gg\mathcal O(\epsilon)$, and varies slowly, $\nu\equiv H^{-1} {\cal D}_t \ln \omega = \mathcal O(\epsilon)$. 
The equations of motion then imply that
\eq{
\dot\phi_v=\frac{-3\Vv}{H(9+\omega^2)},\hspace{1cm}\dot\phi_w=\frac{\omega\Vv}{H(9+\omega^2)},\label{eq:velocities}
}
and it also follows that $\Omega_v/H=\omega$ up to $\mathcal O(\epsilon)$ corrections. The condition that $\nu=\mathcal O(\epsilon)$  imposes that the gradient must vary slowly along the trajectory $H^{-1}\mathcal D_t \ln\Vv=\mathcal O(\epsilon)$. In this solution, the first inflationary slow-roll parameter is given by \cite{Hetz:2016ics}
\beq
 \epsilon = - \frac{\dot H}{H^2} = \frac{1}{(1+ \omega^2/9)}  \frac{V_v^2}{2V^2} \ll 1 \, ,
\eeq
meaning that inflation is possible in steep potentials if $\omega$ is sufficiently large, i.e.~in a rapid-turn solution.

To find explicit rapid-turn solutions we note that the field velocities in equation \eqref{eq:velocities} are given in terms of the gradient of the potential, the Hubble rate, and the turn rate $\omega$. The first two are known immediately at any position in field space, and the latter is fixed by the conditions $\Omega_v/H=\omega$ and $H^{-1}\mathcal D_t\ln\Vv=\mathcal O(\epsilon)$ \cite{Bjorkmo:2019fls}:
\eq{
\frac{\Vww}{H^2}-\frac{9}{\omega^2}\frac{\Vvv}{H^2}=\omega^2+9+\mathcal O(\epsilon),\hspace{1cm}\frac{\Vvw}{H^2}-\frac{3}{\omega}\frac{\Vvv}{H^2}=\mathcal O(\omega\epsilon).
\label{eq:bkgcond}
}
This determines the field velocities as a function of field-space position, much like in slow-roll, slow-turn inflation. Rapid-turn inflation can only arise if these two equations can be simultaneously satisfied, and the precise form of the solution is strongly dependent on the form of the covariant Hessian of the potential projected onto the gradient basis.

It is instructive to concretise these rather abstract considerations by a particular example: hyperinflation \cite{Brown:2017osf}. The hyperbolic plane has a metric that can be written as
\beq
ds^2 = d\varphi^2 +L^2 \sinh^2(\varphi/L) d\theta^2 \, ,
\eeq
where $L$ sets the (Ricci) curvature of the field space: $R=-2/L^2$. For a rotationally symmetric potential, $V=V(\varphi)$, the vielbeins are given by \cite{Bjorkmo:2019aev},
\beq
v^a = (1,0)~~ \text{and }~w^a = \left(0, \tfrac{1}{L\sinh(\varphi/L)} \right) \, .
\eeq
It follows  that $\Vvv=V_{,\varphi\varphi}$, $V_{;vw} =0$ and $V_{;ww} =V_{,\varphi}/L$, which simplifies the conditions \eqref{eq:bkgcond} to
\beq
\omega^2+9=\frac{V_{,\varphi}}{LH^2},\hspace{1cm}\frac{V_{,\varphi\varphi}}{\omega^2 H^2}=\mathcal O(\epsilon)\, .
\eeq
The first equation fixes the turn rate, and the second requires $LV_{,\varphi\varphi}/V_{,\varphi}\ll1$. By noting that $\dot\phi_v=\dot\varphi$ and $\dot\phi_w= L\sinh(\varphi/L)\dot\theta$, one then finds using equation \eqref{eq:velocities} that the field velocities are given by
\eq{
\dot\varphi=-3HL\, ,\hspace{1cm} L\sinh(\varphi/L)\dot\theta=\pm\sqrt{LV_{,\varphi}-9H^2L^2} \, .
}
The assumption of a spherical symmetric potential is not necessary for the hyperinflation solution to exist; as long as the effects of the negative curvature are substantial, hyperinflation can be realised even in steep and random potentials \cite{Bjorkmo:2019aev}.

Hyperinflation and other rapid-turn inflationary models obey a common attractor solution \cite{Bjorkmo:2019fls}. Moreover, a particularly interesting feature of inflationary models in curved field spaces, such as hyperinflation, is that they can be reached from the standard slow-roll solutions through \emph{geometric destabilisation} of slow-roll inflation \cite{Renaux-Petel:2015mga}. In the case of hyperinflation, this happens as soon as $LV_{,\varphi} > 9 H^2L^2$ \cite{Brown:2017osf,Bjorkmo:2019aev}.

\subsection{Linear perturbations}

To study the perturbations it is advantagous to use the kinematic basis
\eq{
	e^I_a=(n_a,~s_a),
}
where $n^a=(\dot\phi_v v^a+\dot\phi_w w^a)/\dot\phi$  and $s^a=(-\dot\phi_w v^a+\dot\phi_v w^a)/\dot\phi$. In the kinematic basis, the effective mass matrix of the perturbation $M_{ab}$ has two out of three independent elements constrained up to $\mathcal O(\epsilon)$ corrections:
\eq{
M_{nn}=\omega^2H^2,\hspace{1cm}M_{ns}=-3\omega H^2 \, .
}
The entropic mass cannot be determined in terms of the turn-rate alone so we write it as
\eq{
M_{ss}\equiv \xi\omega^2 H^2.
}
It will be convenient for us to work with the curvature perturbation $\zeta=n_a\delta\phi^a/\sqrt{2\epsilon}$ and with the entropic perturbation $\sigma=\delta\phi_s$, with which the action takes the following form (see Appendix \ref{app:quadraticaction}):
\eq{
\hspace{-0.15cm}\mathcal S=\frac12\int dt\frac{d^3k}{(2\pi)^3}a^3\left[2\epsilon\left(\dot\zeta^2-\frac{k^2}{a^2}\zeta^2\right)+\dot\sigma^2-\frac{k^2}{a^2}\sigma-H^2\omega^2(\xi-1)\sigma^2-4\sqrt{2\epsilon}\omega H\sigma\dot\zeta\right].  \label{eq:quadact}
}

When $\xi>1$, the perturbations can be described by an effective single-field theory with a reduced speed of sound, which has been studied extensively in the literature. However, when $\xi<1$, the perturbations exhibit interesting behaviours. They have been described by a single-field EFT with an \textit{imaginary} speed of sound  \cite{Garcia-Saenz:2018ifx,Garcia-Saenz:2018vqf,Fumagalli:2019noh}, and these are the ones that we shall focus on in this paper.

 A remarkable feature of theories with $\xi<1$ is that during the last few e-folds before the modes cross the horizon, they undergo a transient instability that causes the power spectrum to grow exponentially. One may certainly fear that such a growth of the perturbations will also be reflected in enhanced non-Gaussianity in higher-order correlation functions, and  we will now review the recent argument by   \cite{Fumagalli:2019noh}, which pointed towards excessive non-Gaussianities from rapidly-turning realisations of hyperinflation, and similar models. 
 
In closing, we note that in section \ref{sec:growth}, we use the WKB method to provide the first analytic expressions for the growing mode functions that apply to the whole class of rapid-turn models with $\xi<1$. Our analysis will 
in particular find good agreement with the single-field EFT with $c_s^2<0$ used in   \cite{Fumagalli:2019noh}, and in the following section we review the argument for suspected exponentially large non-Gaussianities using the simpler EFT description. 

\subsection{Hyper-large non-Gaussianities?}

In the single-field EFT with an imaginary speed of sound, the mode function of the curvature perturbation can be written as
\cite{Garcia-Saenz:2018vqf,Fumagalli:2019noh}
\eq{
	\zeta_k(\tau)=\left(\frac{2\pi^2}{k^3}\right)^{1/2}\alpha\left(e^{k|c_s|\tau+x}(k|c_s|\tau-1)-\rho e^{i\psi}e^{-(k|c_s|\tau+x)}(k|c_s|\tau+1)\right) \, .
	\label{eq:modefuncEFT}
}
The coefficients $\alpha$, $\rho$ and $\psi$ 
are all assumed to have a mild $k$-dependence, although quantisation fixes $\alpha^2\sim H^2/\epsilon\Mp^2$. This EFT is expected to be valid for  $-x/|c_s| < k\tau$. 
Most important here is $x$, which parametrises the magnitude of the power spectrum at horizon crossing (i.e~at the end of the transient growth), and which is expected to be large when the turning rate is large:\footnote{For the more precise relation, see section \ref{sec:growth}.}  $x\sim \omega$  \cite{Fumagalli:2019noh}. 
 The power spectrum at horizon crossing is then given by
 \eq{
	P_\zeta=
	\frac{k^3}{2\pi^2} \langle |\zeta_k|^2 \rangle =
	\alpha^2e^{2x} \, ,
	\label{eq:Pcross}
}
assuming $\rho \lesssim {\cal O}(1)$. As we have $\alpha^2\sim H^2/\epsilon\Mp^2$, it is the factor of $e^{2x}$ that captures the exponential growth of the perturbations. In the full two-field model, the entropic modes decay after horizon crossing and the perturbations become adiabatic and constant, so that equation \eqref{eq:Pcross} gives the final power spectrum that must be matched with observations: $\alpha^2e^{2x} =P_{\rm obs} = 2\times 10^{-9}$. This can be written as a normalisation condition: 
$\alpha\sim 10^{-5} e^{-x}$.

Most interestingly, the EFT treatment of \cite{Fumagalli:2019noh} allowed for the first discussion and calculations of non-Gaussianities in these models. 
Specifically, the bispectrum was found to peak for flattened configurations (for which $ k_2=  k_3=k_1/2$) with an amplitude of $f^{\rm flat}_{\rm NL} = {\cal O}(50)$ for one example of hyperinflation. Most importantly however,  some contributions to the (non-Gaussian) connected $n$-point correlation functions for $n\geq4$ were analytically found to be \emph{exponentially large}, leading to an apparent loss of perturbative control. 

Key to this discussion is the expansion of the curvature perturbation $\zeta$ in terms of a Gaussian field $\zeta_g$:
\eq{
\zeta=\zeta_g\left(1+f^{(1)}_\text{NL}\zeta_g+f_\text{NL}^{(2)}\zeta_g^2+\dots\right),
}
where the coefficients $f^{(n-2)}_\text{NL}$ are given by
\eq{
f^{(n-2)}_\text{NL}=\frac{\langle \zeta^{n}\rangle_c}{\langle\zeta^{2n-2}\rangle},
}
and the subscript $c$ denotes a connected correlation function. For the expansion to be well-defined, we require (heuristically)
\eq{
f^{(n-2)}_\text{NL}|\zeta_g|^{n-2}\sim f^{(n-2)}_\text{NL}\langle\zeta^2\rangle^{(n-2)/2}\lesssim1.
}
Thus, as argued in Reference \cite{Fumagalli:2019noh}, we retain perturbative control as long as 
\eq{
	\frac{\langle\zeta^n\rangle_c}{\langle \zeta^2\rangle^{n-1}}\lesssim\langle\zeta^2\rangle^{-(n-2)/2} \, .
\label{eq:cond}
}

Some contributions to $\langle \zeta^n\rangle_c$ were shown to be harmless in \cite{Fumagalli:2019noh}: contact interactions lead to no exponential enhancement in the non-Gaussian parameters. This explains why, in particular, the connected three-point function is not very large. However, other contributions that involve the tree-level exchange of (a scalar) $\zeta$, were found to be dangerous. 
The starting point for this argument is the expansion of the correlators in the in-in formalism
\eq{
	\langle\hat\zeta^n(\tau)\rangle=\sum_{k=0}^\infty i^k\int_{-\infty}^\tau d\tau_1\ldots\int_{-\infty}^{\tau_{k-1}}d\tau_k\langle[\Hint(\tau_k),\ldots [\Hint(\tau_1),\hat\zeta^n(\tau)]\ldots]\rangle,
}
where the operators on the right-hand side are in the interaction picture. To evaluate these correlators, one uses the mode functions of the (free) quadratic theory, and incorporates cubic and higher-order interaction terms through $\Hint$. 

\begin{figure}
    \centering
    \includegraphics[width=0.5\textwidth]{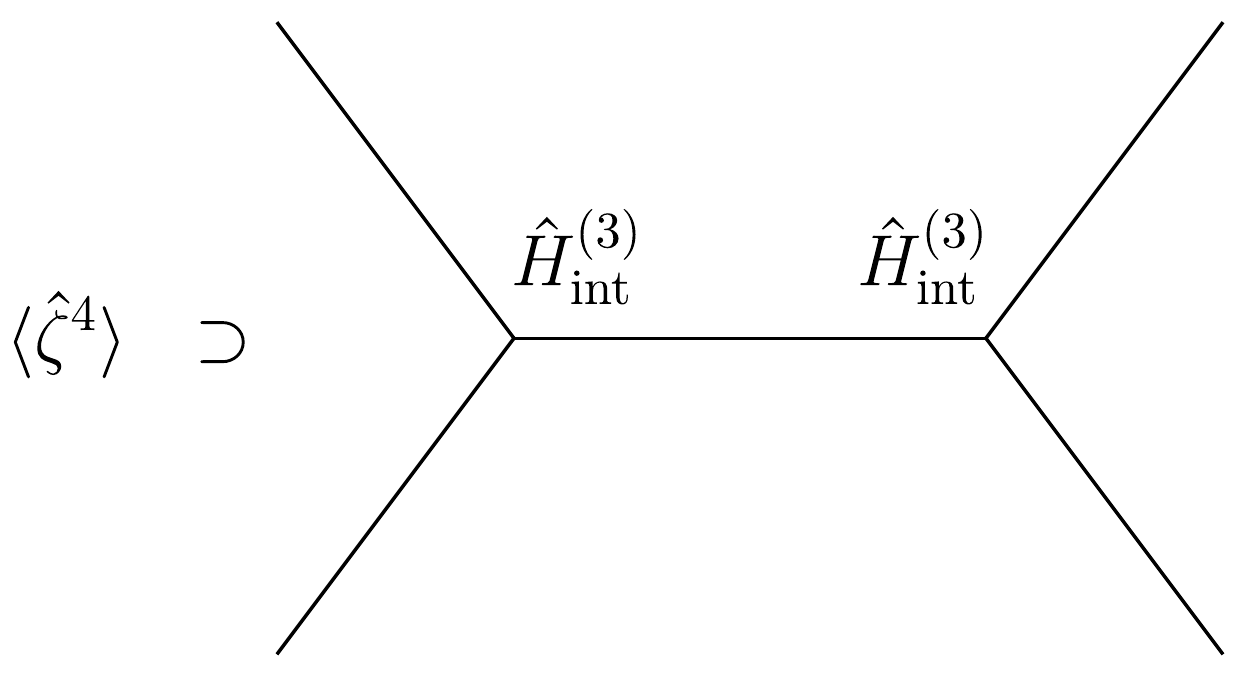}
    \caption{The four-point correlation function gets a contribution from a scalar exchange diagram, corresponding to two insertions of the cubic interaction Hamiltonian. \label{fig:4ptX}}.
\end{figure}

The simplest example of an arguably dangerous contribution to the four-point function is show in Figure \ref{fig:4ptX}. The corresponding contribution to the connected 4-point function is in the in-in formalism given by
\eq{ \label{eq:zeta4pf}
	\langle \hat\zeta^4(\tau)\rangle_c\supset -\int_{-\infty}^\tau d\tau_1 \int_{-\infty}^{\tau_1}d\tau_2 \langle[\Hint(\tau_2),[\Hint(\tau_1),\hat\zeta^4(\tau)]]\rangle,
}
with cubic interactions $\Hint\sim \tfrac{1}{\alpha^2} \zeta^3$. The actual interactions involve derivatives of $\zeta$, but these do not affect the exponential scaling or the general argument. Moreover, the important factor of $\alpha^{-2}$ comes from the factor of $\epsilon\Mp^2/H^2$ in the cubic vertex. We will explain both these points in more detail in section \ref{sec:npoint}.

The integrand of this contribution to \eqref{eq:zeta4pf} then involves 10 powers of $\hat \zeta$ (three respectively at $\tau_1$ and $\tau_2$, and four at $\tau$), which should all be appropriately contracted using Wick's theorem. The naive scaling of this diagram is therefore:
\beq
\langle \hat\zeta^4(\tau)\rangle_c \sim \alpha^{6} e^{10x} 
\hspace{80 pt}(\text{naive})
\label{eq:4ptnaive}
\eeq
It then follows from \eqref{eq:Pcross} that the ratio \eqref{eq:cond} scales like
\beq
g_{\rm NL} =\frac{\langle\zeta^4\rangle_c}{\langle \zeta^2\rangle^{3}} \sim e^{4x} \, , \hspace{80 pt}(\text{naive}) 
\label{eq:gnlnaive}
\eeq
which is exponentially large and will even be larger than $P_{\zeta}^{-1}$ for $x\gtrsim5$.

Thus, it would seem that hyperinflation and other rapidly turning models are severely constrained by limits on non-Gaussianity, currently at the level $g_{\rm NL} \lesssim {\cal O}(10^{4}-10^{6})$ \cite{Akrami:2019izv, Meerburg:2019qqi}, and even the requirement of perturbativity. However, as we will now show, such a conclusion would  be too quick and, in fact, inaccurate.
The estimates 
 \eqref{eq:4ptnaive} and \eqref{eq:gnlnaive}
 do not account for subtle yet exact cancellations within the relevant correlation functions, and the actual non-Gaussian parameters of these models  is merely ${\cal O}(1)$ --- far from observational constraints or tensions with perturbativity. 

\section{Perturbativity and $n$-point functions \label{sec:npoint}}

Reference \cite{Fumagalli:2019noh} found that while 4-point and higher-order correlation functions see an exponential amplification of non-Gaussianity due to the growth of the perturbations around horizon crossing, the bispectra of rapid-turn models were consistent with current observational bounds due to interference between exponentially growing and decaying modes. In this section, we show that this is in fact also the case for the trispectra and higher-order correlation functions. The terms which one naively would expect to be dominant instead cancel out, and the dominant terms will be products of exponentially growing and decaying modes, limiting the overall growth. These results are analagous to those for gauge fields axially coupled to the inflaton during inflation \cite{Ferreira:2015omg}.

%

\subsection{The 4-point correlator}
Before we consider the general $n$-point correlator, we will for simplicitly consider the specific case of the 4-point correlation function, and show that $g_\text{NL}$ is not outside of observational bounds. We will in particular show show why the term in the correlator arising from two insertions of cubic Hamiltonians must be proportional to $e^{6x}$ (instead of $e^{10x}$ as in equation \eqref{eq:4ptnaive}), and gives $g_\text{NL}\sim1$ (instead of $g_{\rm NL} \sim e^{4x}$ as in equation \eqref{eq:gnlnaive}).

The correlator inside the integral in equation \eqref{eq:zeta4pf} can be expanded as
\begin{align*}
	\langle[\Hint(\tau_1),[\Hint(\tau_2),\hat\zeta^4(\tau)]]\rangle&= \langle \Hint(\tau_1)\Hint(\tau_2)\hat\zeta^4(\tau)\rangle- \langle \Hint(\tau_1)\hat\zeta^4(\tau)\Hint(\tau_2)\rangle\\
	&~~~+\langle \hat\zeta^4(\tau)\Hint(\tau_2)\Hint(\tau_1)\rangle - \langle \Hint(\tau_2)\hat\zeta^4(\tau)\Hint(\tau_1)\rangle\\
	&=2\,\mathrm{Re}\left(\langle \Hint(\tau_1)\Hint(\tau_2)\hat\zeta^4(\tau)\rangle\right)\\
	&~~~-2\,\mathrm{Re}\left(\langle \Hint(\tau_1)\hat\zeta^4(\tau)\Hint(\tau_2)\rangle\right)\nt
\end{align*}
where in the second line we used the fact that the Hamiltonians and observable n-point correlators are Hermitian. Within each of these correlators, we now have Wick contractions turning them into all products of five pairs of  $\zeta_1\zeta_2^*$. Writing the generic mode functions from the single-field EFT
as\footnote{We have chosen this form for the mode functions in order to make the argument more transparent. Adding $k$-dependent phases to any of the two terms in this equation does not change the results derived in this work. We have not however considered the possibility that corrections in the form of time-dependent phases could show up in one of the two terms. Although unlikely we cannot exclude that such terms could show up in loop corrections to $\zeta$.} (cf.~equation \eqref{eq:modefuncEFT})
\eq{ \label{zetamodefunction}
	\zeta_i(\tau)=f_i(\tau) e^x+ig_i(\tau)e^{-x},
}
we see that the terms proportional to $e^{10x}$ in \textit{both} terms will be
\eq{
	e^{10x}\prod_i f_i,
}
and they will hence cancel out after we contract all combinations. Any term proportional to $e^{8x}$ must be imaginary, and hence does not contribute either. Only at $e^{6x}$ do we expect the terms not to cancel out. Another way to see this is to note that a Wick contraction between real mode functions is a symmetric operator. Since the $e^{10x}$ term comes from only the real part of the mode functions, the order of the operators inside the correlator does not matter and the two terms will cancel out.

To make the discussion a bit more precise, the interaction Hamiltonian contains terms of the form \cite{Garcia-Saenz:2018vqf,Fumagalli:2019noh,Garcia-Saenz:2019njm}
\begin{align*}
	\Hint&\sim\int d^3x\frac{a\epsilon\Mp^2}{H}\left(\frac1{c_s^2}-1\right) (\partial \hat\zeta)^3\sim\int d^3x\frac{-1}{\tau}\frac{\epsilon\Mp^2}{H^2}\left(\frac1{c_s^2}-1\right)(\partial \hat\zeta)^3\\
	&\sim\int d^3x\frac{-1}{\tau}\alpha^{-2}\left(\frac1{c_s^2}-1\right)(\partial \hat\zeta)^3.
\end{align*}
 The conservation of $\zeta$ on superhorizon scales then implies that the time integrals in equation \eqref{eq:zeta4pf} peak around the horizon crossing time of the modes. The presence of derivatives does affect the argument so we can neglect them. Therefore, the contribution to $\langle \zeta^4\rangle$ from two insertions of this operator then scales as
\eq{ \label{eq:4pfscaling}
	\left(\frac1{c_s^2}-1\right)^2\alpha^{6}e^{6x}
}
and it follows that $g_\text{NL}$ will satisfy
\eq{
	g_\text{NL}\sim\frac{\langle\zeta^4\rangle}{\langle \zeta^2\rangle^3}\sim\left(\frac1{c_s^2}-1\right)^2 \frac{\alpha^{6}e^{6x}}{\alpha^6e^{6x}}\sim1.
}
At $n=4$, we retain perturbative control as long as
\eq{
	g_\text{NL}\lesssim P_\zeta^{-1},
}
which is satisfied with a very good margin for theories with $P_\zeta\sim 10^{-9}$.

As an aside, we note that this apparent suppression is not surprising as it is common in situations where both the initial and the final states have large occupation numbers $N(k)$. 
For example, in a 2 to 2 scattering, if the phase space density of each state $i$ (in or out) has  $N_i(k_i) \gg 1$ there is a Bose enhancement of the process proportional to
\begin{eqnarray}
N_1(k_1) N_2(k_2)[1+N_3(k_3)][1+N_4(k_4)]-N_3(k_3) N_4(k_4)[1+N_1(k_1)][1+N_2(k_2)] \propto N_i^3. \quad \, \,
\end{eqnarray}
where the first term is associated with the process $1+2 \rightarrow 3+4$ and the second with $ 3+4 \rightarrow 1+2 $. The leading term, proportional to $N_i^4$, is canceled between those two processes. If we then interpret the 4-point function in equation \eqref{eq:zeta4pf} as a 2 to 2 scattering we see that the final $N_i^3$ scaling matches with the result in equation \eqref{eq:4pfscaling} where the phase space density is nothing more than $N_i(k/a\simeq H) \propto |\zeta|^2 \propto e^{2x}$.

\subsection{The general $n$-point correlator}

To show that the same holds for all higher $n$-point correlation functions, we need to take a different approach, as the previous one does not generalise easily to higher orders.

As a starting point, we consider the commutator of some (at least cubic) interaction Hamiltonian $\Hint$ with some product of operators $\hat A$:
\eq{
	\langle[\Hint,\hat A]\rangle=\langle \Hint \hat A\rangle-\langle \hat A\Hint\rangle.
}
If $\Hint$ contains an odd number of operators, at least one of them must be contracted with some operator(s) in $\hat A$. If $\Hint$ contains an even number of operators, the terms with all operators contracted internally will cancel out (see RHS), and hence at least two must be contracted with operators in $\hat A$. Therefore, any surviving terms will have contractions between some of the operators in $\Hint$ and some of those in $\hat A$.

This has important consequences. As we shall see below, this always gives the imaginary part of the products of the mode functions, and we will show that the only non-zero terms from the expectation of nested commutators involve contractions between terms on the LHS and RHS of each commutator. Every time we do this, we will pick up a factor of the imaginary part of the product of some number of mode functions. Since the imaginary parts of the mode functions scale as $\alpha e^{-x}$, this limits the size of the connected $n$-point correlators arising from the cubic scalar exchange vertex.

To see why we get the imaginary part of the product of the mode functions, consider the terms containing contractions of some operators $\hat \zeta_{1a},\hat \zeta_{1b}...$ on the left hand side of the commutator with $\hat \zeta_{2a},\hat \zeta_{2b}...$ on the right hand side. These operators are either the curvature perturbation or various derivatives of it, but have the same annihilation and creation operators. We thus have
\begin{align*}
	&\wick{\langle\dots [\dots \c1 {\zeta}_{1a}\dots \c2{\zeta}_{1b}\dots,\dots \c2{\zeta}_{2b}\dots \c1{\zeta}_{2a}\dots]\dots\rangle}\\
	&\propto\zeta_{1a}\zeta_{2a}^*\zeta_{1b}\zeta_{2b}^* \ldots - \zeta_{2a}\zeta_{1a}^*\zeta_{2b}\zeta_{1b}^* \ldots = 2i\,\mathrm{Im}\left(\zeta_{1a}\zeta_{2a}^*\zeta_{1b}\zeta_{2b}^*\dots\right),
\end{align*}
which will give us terms where we pick up an odd number of imaginary parts of the mode functions. It is important to note that these contractions do not depend on the relative positions of the operators $\hat \zeta_{2a},\hat  \zeta_{2b}...$ on the RHS. It does not matter if commutators inside the RHS shuffle these around or if some other operators inside the commutator are contracted with operators outside of it -- it always gives the same factor. Therefore, terms with Wick contractions across $n$ commutators, whether they are nested or not, are proportional to the product of $n$ factors of imaginary parts of products of mode functions.

The mode functions in the EFT we consider can be written as in equation \eqref{zetamodefunction}.
Hence, the dominant term in  $\mathrm{Im}(\zeta_1\dots\zeta_n^*)$ does not scale as $e^{nx}$, as one might naively expect, but instead as
\eq{
	\mathrm{Im}(\zeta_1\dots\zeta_n^*)\propto e^{(n-1)x}e^{-x}=e^{(n-2)x}.
}
In fact, every time we have a commutator we will see this relative suppression of the expectation value by $e^{-2x}$ compared to the naively expected one, drastically limiting the size of certain diagrams. The reason for this is that when we have nested commutators all non-zero terms will have contractions across each commutator, giving us factors of imaginary parts of products of mode functions. We will argue why this is the case below.

To proceed, we want to consider a nested commutator with operators $\hat H_1\equiv \Hint(\tau_1)$, $\hat H_2\equiv \Hint(\tau_2)$ and so forth. We begin by looking at the case with two nested commutators:
\eq{
	\langle[\hat H_1,[\hat H_2,\hat A]]\rangle.
}
As we saw above, non-zero terms must have contractions of operators in $\hat H_1$ with operators in $[\hat H_2,\hat A]$. Expanding the above, we have
\eq{
	\langle[\hat H_1,[\hat H_2,\hat A]]\rangle=\langle \hat H_1[\hat H_2,\hat A]\rangle-\langle[\hat H_2,\hat A]\hat H_1\rangle.
}
The crucial point is that terms in $\langle \hat H_1[\hat H_2,\hat A]\rangle$ or $\langle[\hat H_2,\hat A]\hat H_1\rangle$ with no operators in $\hat H_2$ contracted with any in $\hat A$ must necessarily vanish. To see this, we note that for these terms, operators in $\hat A$ are contracted either internally or with those in $\hat H_1$, and the same goes for $\hat H_2$. Terms with all operators contracted internally within $\hat A$ vanish due to the commutator, hence at least some operators must be contracted with those in $\hat H_1$. But these terms vanish too, because (schematically)
\eq{
	\wick{\langle \c1 H_1[H_2, \c1 A]\rangle}=\wick{\langle \c1 H_1H_2 \c1 A\rangle}-\wick{\langle \c1 H_1 \c1 A H_2\rangle}=0,
}
as a consequence of no operators in $\hat A$ being contracted with any in $\hat H_2$. None of the terms above depend on the relative position of $\hat H_2$ and $\hat A$, and so they cancel out exactly. It is therefore only if some operators in $\hat H_2$ are contracted with operators in $\hat A$ that these expressions can be non-zero.

We are now free to replace $\hat A$ with $[\hat H_3,\hat B]$, giving
\begin{align*}
	\langle[\hat H_1,[\hat H_2,[\hat H_3,\hat B]]]\rangle&=\langle \hat H_1\hat H_2[\hat H_3,\hat B]\rangle-\langle \hat H_1[\hat H_3,\hat B]\hat H_2\rangle\\
	&~~~-\langle \hat H_2[\hat H_3,\hat B]\hat H_1\rangle+\langle [\hat H_3,\hat B]\hat H_2\hat H_1\rangle,\nt
\end{align*}
and again, each term will vanish unless some operators in $\hat H_3$ are contracted with some in $\hat B$. We can repeat this argument indefinitely, and the result can be summarised as: \textit{In evaluating the expectation of $n$ nested commutators, all non-zero terms will include at least one operator on the LHS of each commutator contracted with operators on the RHS.}

Putting it all together, we therefore see that with $n-2$ insertions of a cubic interaction Hamiltonian $\Hint$, the dominant contribution to the $n$-point correlator therefore scales as
\begin{align*}
	\langle[\Hint(\tau_1),\dots,[\Hint (\tau_{n-2}),\hat\zeta^n]\dots]\rangle&\propto\alpha^{-2(n-2)} \alpha^{4n-6}e^{(4n-6)x}e^{-2(n-2)x}\\
	&\propto\alpha^{2n-2}e^{(2n-2)x}
\end{align*}
instead of the naive $\alpha^{2n-2}e^{(4n-6)x}$. We then find
\eq{
	\frac{\langle \zeta^n\rangle}{\langle \zeta^2\rangle^{n-1}}\sim\frac{\alpha^{2n-2}e^{(2n-2)x}}{\alpha^{2n-2}e^{(2n-2)x}} =1,
}
causing no issues with perturbative control.

%

Do we expect these result hold even when we include loop corrections? Yes, because for every insertion of the cubic interaction Hamiltonian we pick an overall factor of $\alpha e^{x}=P_\zeta^{1/2}\ll1$. The more of these we insert (and we need two for every loop correction), the greater the suppression is (prior to integration). There is therefore a priori no reason to expect that loop corrections will change these results.

\section{Linear perturbations in rapid-turn inflation \label{sec:growth}}

To understand the exponential amplification of the curvature perturbation in theories with $\xi<1$ we now look at the quadratic action for the full two-field theory. Here the aim is to compute the mode function of $\zeta$ analytically (neglecting Hubble friction, similarly to references \cite{Achucarro:2012yr,Cespedes:2012hu}) using a WKB approach, and thus compute the growth of the power spectrum in general. It will also provide further evidence for the validity of the EFT. The equations of motion resulting from equation \eqref{eq:quadact} are
\begin{align}
\ddot\zeta+3H\dot\zeta+\frac{k^2}{a^2}\zeta&=\frac{2\omega H}{\sqrt{2\epsilon}}(\dot\sigma+3H\sigma)\\
\ddot\sigma+3H\dot\sigma+\frac{k^2}{a^2}\sigma+H^2\omega^2(\xi-1)\sigma^2&=-{2\omega H}{\sqrt{2\epsilon}}\dot\zeta,
\end{align}
and the goal of this section is to find and understand their solutions. To make progress we assume that we are in a phase where Hubble friction can be neglected (which given the exponential growth and $\omega \gg 1$ is a fair assumption), and make the ansatz
\eq{
\zeta=\zeta_+e^{i\lambda_+t}+\zeta_-e^{i\lambda_-t},\hspace{1cm} \sigma=\sigma_+e^{i\lambda_+t}+\sigma_-e^{i\lambda_-t}.
}
Both of these frequencies should of course appear twice, with different signs, but for notational convenience we ignore this at the moment. We then find that the frequencies $\lambda_\pm$ are given by
\eq{
\lambda_\pm^2/H^2\equiv \tilde\lambda_\pm^2=\kappa^2+\frac{3+\xi}2\omega^2\pm\frac12\sqrt{16\kappa^2\omega^2+(3+\xi)^2\omega^4},
}
where we have introduced the notation $\kappa\equiv k/aH$. It will shortly be useful for us to work with e-folds as a time coordinate, and when ignoring $\mathcal O(\epsilon)$ corrections we are free to shift it (for each $k$-mode) such that $\kappa=e^{-N}$, meaning that horizon exit happens at $N=0$.


The first thing we note is that for $\xi<1$, $\lambda_-$ becomes imaginary for sufficiently small $\kappa$. This happens at
\eq{
\kappa^2+\frac{3+\xi}2\omega^2=\frac12\sqrt{16\kappa^2\omega^2+(3+\xi)^2\omega^4}\hspace{0.5cm} \Rightarrow\hspace{0.5cm}\kappa=\sqrt{1-\xi}\omega,
}
and therefore from $N=-\ln(\sqrt{1-\xi}\omega)$ e-folds before horizon crossing and onwards, $\lambda_-$ is imaginary. This means that during this phase the mode functions will grow exponentially, which is exactly what one finds numerically. In hyperinflation, $\xi=-1$, and we recover Brown's result that the growth starts at $\ln(\sqrt{2}\omega)$ e-folds before horizon crossing \cite{Brown:2017osf}.

We are now in a position to compute an analytic approximation for the mode functions for $\zeta$ without an EFT. This can be done by integrating $|\tilde\lambda_-|$ from $N=-\ln(\sqrt{1-\xi}\omega)$ to up to some arbitrary $N$ (in effect, we are using the WKB method), and it is more accurate than one might initially expect, since the contributions from Hubble friction are negligible on superhorizon scales. Here we let $\zeta_\pm$ refer to the positive and negative frequency solutions of the low frequency modes, and they can be written as
\eq{
\zeta_\pm\propto\exp\left[\pm \mathcal I(N)\right],
}
where the integral $\mathcal I(N)$ is given by 
\eq{
\mathcal I(N)=\int_{-\ln(\sqrt{1-\xi}\omega)}^N\sqrt{-\kappa^2-\frac{3+\xi}2\omega^2 +\frac12\sqrt{16\kappa^2\omega^2+(3+\xi)^2\omega^4}}dN.
}
One can show after tedious algebra (see Appendix \ref{app:growth}) that this integral can be evaluated to
\eq{
\mathcal I(N)=F(b-1)-F(\sqrt{1+16\kappa^2/(3+\xi)^2\omega^2}),
}
where the function $F$ is given by,
\begin{align*}
F(y)&=\frac{2\omega}b\sqrt{(y-1)(b-1-y)}- \omega \arctan\left[\frac{b-2y}{2\sqrt{(y-1)(b-1-y)}}\right]\\
&\hspace{0.5cm}-\omega\sqrt{\frac 2b}\arctan\left[\frac{\sqrt2 (2-3b+2y+by)}{4\sqrt{b(y-1)(b-1-y)}}\right]\nt,
\end{align*}
and we have defined $b=8/(3+\xi)$. While this is not an easy expression to work with, it is  accurate. Of particular interest to us is the behaviour of the mode functions in the regime of validity of the EFT, when $\kappa^2/\omega^2$ is small. Here we find that they simplify to
\eq{
\zeta_\pm\propto\exp\left[\pm(2-\sqrt{3+\xi})\frac{\pi\omega}2\mp|c_s|\kappa\right], \label{eq:zetaLateModeFunctions}
}
where $|c_s|=\sqrt{(1-\xi)/(3+\xi)}$, recovering the leading order behaviour of the $\zeta$ mode function in the EFT \cite{Garcia-Saenz:2018vqf,Fumagalli:2019noh}. The result is precisely what one obtains in the EFT if Hubble friction is ignored. Now, however, we can identify the previously unknown quantity $x$:
\eq{
x(\omega,\xi)=\left(2-\sqrt{3+\xi}\right)\frac{\pi\omega}2.
}

With the above expressions, we can also give an analytic expression for the growth of the power spectrum before horizon crossing, denoted by $\gamma^2=\gamma^2(\omega,\xi)=P_\zeta(\omega,\xi)/P_\zeta(0)$. We assume that once we are on superhorizon scales, $\zeta_+$ is dominant, and that $\zeta_+$ and $\zeta_-$ had roughly equal power at $N=-\ln(\sqrt{1-\xi}\omega)$. Then, for consistency also neglecting the Hubble friction for the single-field $\zeta$, we find that the relative growth of the power spectrum is given by
\eq{
\ln(\gamma^2)\approx(2-\sqrt{3+\xi})\pi\omega,\label{eq:Pzetagrowth}
}
which is obtained by letting $\kappa\to0$ in equation \eqref{eq:zetaLateModeFunctions}. In hyperinflation, we then find that $\ln(\gamma)\propto0.920\,\omega$, similar to the numerical result of Mizuno et al. that $\ln(\gamma)\propto0.924\,\omega$ \cite{Mizuno:2017idt}.

The expression for the $\zeta$ mode function we derived here is an approximation, but it is remarkably accurate. As shown in Figure \ref{fig:lngamma}, the formula for the growth of the power spectrum given in equation \eqref{eq:Pzetagrowth} agrees very well with numerics. 

\begin{figure}
    \centering
    \begin{subfigure}{0.485\textwidth}
    \centering
    \includegraphics[width=1\textwidth]{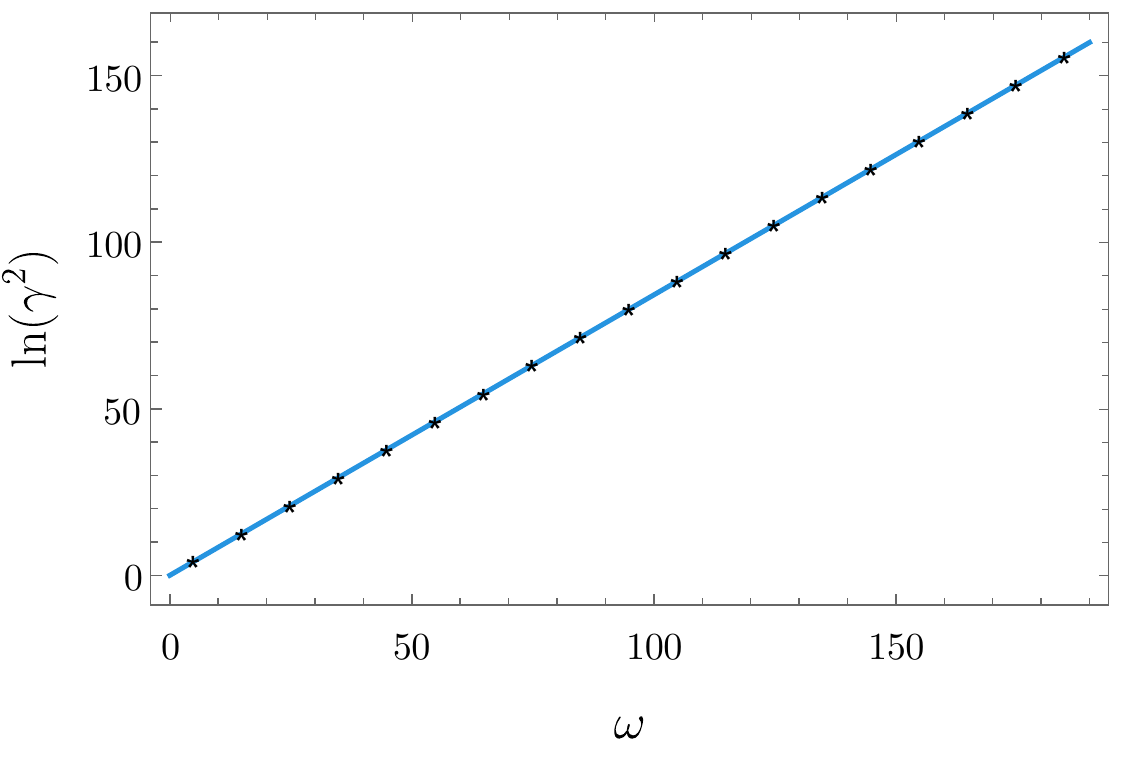}

    \end{subfigure}
    ~ 
    \begin{subfigure}{0.485\textwidth}
         \includegraphics[width=1\textwidth]{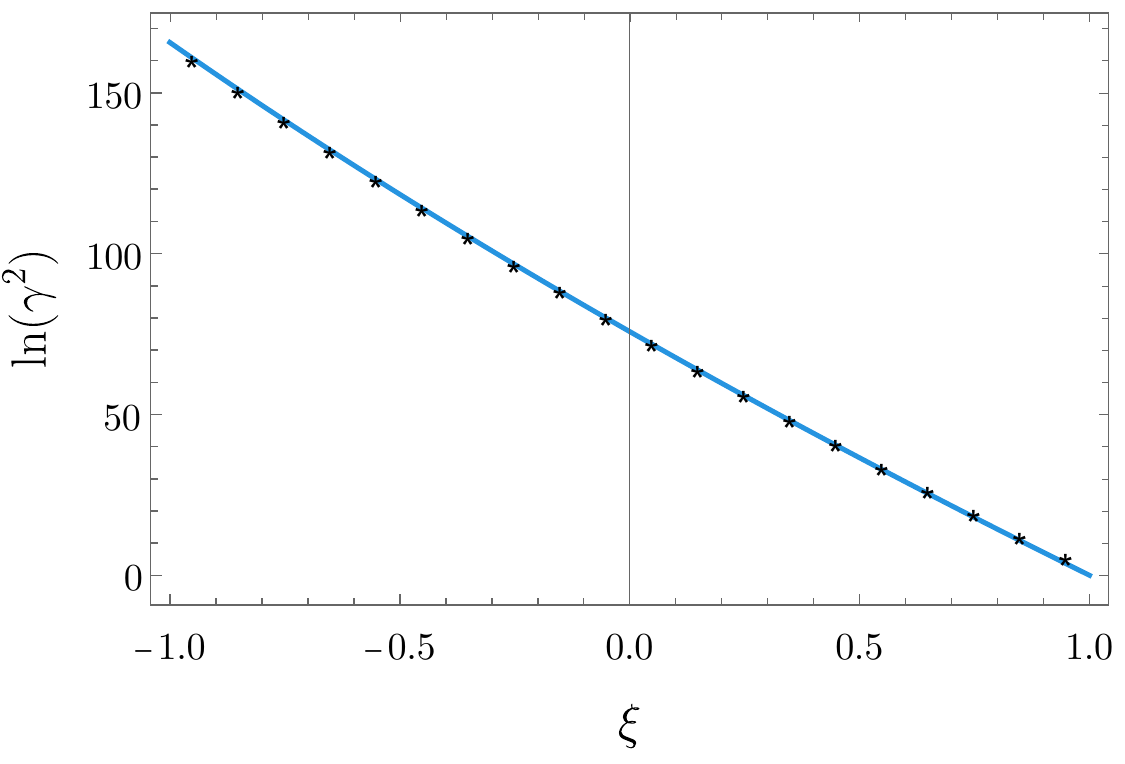}
    \end{subfigure}
    \caption{Comparison between predicted growth and numerically calculated values. The graph on the left has $\xi=0$ fixed and the one on the right has $\omega=90$.\label{fig:lngamma}}.
\end{figure}

Equation \eqref{eq:Pzetagrowth} allows us to determine the maximal turn rate that can be realised in an observationally compatible inflationary model. Sometime after inflation has ended, the universe is reheated to create the conditions for the hot Big Bang cosmology. To ensure the success of Big Bang Nucleosynthesis (and the thermalisation of the neutrinos), reheating must happen at $T_{\rm reheat} > T_{\rm min} \approx4$ MeV. Enforcing (very conservatively) that the Hubble parameter at horizon crossing, $H_*$, is larger than the minimal value $H_{\rm min}$, that $\epsilon_*<1$, and that $P_{\zeta}$ has the correct amplitude then gives
\beq
\ln \gamma^2 \lesssim \ln \left(8\pi^2 P_\text{obs} \left( \frac{M_{\rm Pl}}{T_{\rm min}}\right)^4 \right)  \approx 176 \, .
\eeq
This bound then constrains $\omega$ and $\xi$ through equation \eqref{eq:Pzetagrowth}. For example, in hyperinflation with $\xi=-1$ the turn rate is bounded by $\omega \lesssim 96$.

\section{Conclusions}

In this paper we have further developed the perturbation theory of rapid-turn inflation, which is particularly motivated  by models with negatively curved field spaces. In this class of theories, the curvature perturbation undergoes a transient, exponential amplification before horizon crossing. It has previously been argued, using a single-field EFT, that this amplification also leads to a severe amplification of non-Gaussianity in the trispectrum and higher-order correlation functions. However, as we have shown in this paper, due to the nested structure of the commutators in the in-in formalism correlation functions, the (naive) leading-order terms cancel out exactly, and the remaining non-Gaussianities are mild, in no tension with current observational bounds. These results are similar to the case where gauge fields are exponentially enhanced during inflation \cite{Ferreira:2015omg}.

We have also provided an analytic expression for the scalar curvature perturbation in the full two-field theory, which we found to be in good agreement with the one derived in the single-field EFT with an imaginary speed of sound. This accurate approximation allowed us to find an analytic expression for the total growth of the power spectrum during the transient instability, which closely matches numerical results.

Our results vindicate rapid-turn inflation in hyperbolic field spaces as a viable model for the early universe, and 
confirm the applicability of
the single-field EFT with an imaginary speed of sound as a valid and accurate description of the primordial perturbations. 

\subsection*{Acknowledgements}
We thank Anne-Christine Davis for stimulating discussions, and we are very grateful to Sebastien Renaux-Petel for encouraging discussions and comments on a draft of this paper.
T.B. is funded by an STFC studentship at DAMTP, University of Cambridge, and is grateful to the organisers of
the `Inflation and Geometry'  workshop in June 2019 at the IAP for their kind hospitality.
 D.M. acknowledges support from the Swedish Science Agency grant 2018-03641 and from European Research Council Grant 742104.
\newpage

\appendix

\section{Quadratic action \label{app:quadraticaction}}

The perturbations $\delta\phi^I=e^I_a\delta\phi^a$, are governed by the action \cite{GrootNibbelink:2001qt,Gordon:2000hv,Sasaki:1995aw,Achucarro:2010jv,Achucarro:2010da,Langlois:2008mn}:
\eq{
\mathcal S_{(2)}=\frac12\int\frac{d^3k}{(2\pi)^3}dta^3\left[\delta_{IJ}\mathcal D_t\delta\phi^I_{\vv k}\mathcal D_t\delta\phi^J_{-\vv k}-\left(\frac{k^2}{a^2}\delta_{IJ}+M_{IJ}\right)\delta\phi^I_{\vv k}\delta\phi^J_{-\vv k}\right] ,\label{eq:IntroQuadS}
}
where the covariant derivatives now act through $\mathcal D_t X^I=\dot X^I+\tensor Y{^I_J}X^J$, with $\tensor Y{^I_J}\equiv e^I_a\mathcal D_te_J^a$. In the kinematic basis, this rotation matric $\tensor Y{^I_J}$ is given by
\eq{
\tensor Y{^I_J}=H\begin{pmatrix}0 & -\omega\\ \omega & 0\end{pmatrix}.
}
$M_{IJ}$ is the effective mass matrix, given by
\eq{
M_{IJ}=V_{;IJ}-R_{IKLJ}\dot\phi^K\dot\phi^L+(3-\epsilon)\frac{\dot\phi_I\dot\phi_J}{\Mp^2}+\frac{\dot\phi_IV_{,J}+V_{,I}\dot\phi_J}{H\Mp^2},
}
where $V_{;ab}$ is the second covariant derivative of the potential on the target space, and all quantities above have been projected onto the vielbein basis.

In rapid-turn inflation, the evolution equations for the perturbations are completely determined, up to $\mathcal O(\epsilon)$ corrections, by the turn rate, the entropic mass, and the Hubble rate.  One can immediately show that two of the three independent elements of the mass matrix are given by 
\eq{
\frac{n^an^b M_{ab}}{H^2}=\omega^2+\mathcal O(\epsilon),\hspace{1cm}\frac{n^as^b M_{ab}}{H^2}=-3\omega+\mathcal O(\omega\epsilon).
}
The entropic mass is the only onconstrained degree of freedom in the mass matrix, and it is given by
\eq{
\frac{s^as^b M_{ab}}{H^2}=\frac{9 \Vww }{H^2(\omega^2+9)}+\frac{\Vvv }{H^2}+\frac{R\dot\phi^2}{2H^2}+\mathcal O(\epsilon)\equiv\xi\omega^2\label{eq:tMss}.
}

\section{Approximate growth of the eigenmode \label{app:growth}}

The integral that computes the $\zeta$ mode function, and consequently the approximate growth of the power spectrum, in rapid-turn inflation with $\xi<1$, is given by
\eq{
\mathcal I(N)=\int_{-\ln(\sqrt{1-\xi}\omega)}^N\sqrt{-\kappa^2-\frac{3+\xi}2\omega^2 +\frac12\sqrt{16\kappa^2\omega^2+(3+\xi)^2\omega^4}}dN.
}
where $\kappa=e^{-N}$. We now define
\eq{
(3+\xi)^2\omega^4+16\omega^2\kappa^2\equiv\omega^4 (3+\xi)^2y^2\hspace{0.5cm}\Leftrightarrow\hspace{0.5cm} y^2=1+\frac{16\kappa^2}{(3+\xi)^2\omega^2},
}
giving (after changing integration limits)
\eq{
\mathcal I(N)=\frac{2\omega}{b}\int_{\sqrt{1+\frac{16\kappa^2}{\omega^2(3+\xi)^2}}}^{b-1} \left(b(y-1)-(y^2-1)\right)^{1/2}\frac{ydy}{y^2-1},
}
with $b\equiv8/(3+\xi)$. One can show that the primitive function of the above is given by
\begin{align*}
F(y)&=\frac{2\omega}b\sqrt{(y-1)(b-1-y)}- \omega \arctan\left[\frac{b-2y}{2\sqrt{(y-1)(b-1-y)}}\right]\\
&\hspace{0.5cm}-\omega\sqrt{\frac 2b}\arctan\left[\frac{\sqrt2 (2-3b+2y+by)}{4\sqrt{b(y-1)(b-1-y)}}\right]\nt,
\end{align*}
and hence the integral is given by
\eq{
\mathcal I(N)=F(b-1)-F(\sqrt{1+16\kappa^2/(3+\xi)^2\omega^2}).
}

To compute the overall growth, we need to evaluate $F(1)$ and $F(b-1)$. The first term contributes nothing in this limit, but the other two terms are somewhat non-trivial. Both the arguments of the arctan functions diverge in these limits, but noting that since $\xi<1$ implies $b>2$, one finds (with $y=1+\delta$ and $y=b-1-\delta$ respectively)
\begin{align*}
F(1)&=\lim_{\delta\to0^+}\left(-\omega\arctan\left[\frac{\sqrt{b-2}}{2\sqrt{\delta}}\right] -\omega\sqrt{\frac2b}\arctan\left[\frac{-\sqrt{b-2}}{\sqrt{2b\delta}}\right]\right)\\
&=-\frac{\pi\omega}2+\sqrt{\frac2b}\frac{\pi\omega}2=-(2-\sqrt{3+\xi})\frac{\pi\omega}4\nt\\
F(b-1)&=\lim_{\delta\to0^+}\left(-\omega\arctan\left[\frac{-\sqrt{b-2}}{2\sqrt{\delta}}\right] -\omega\sqrt{\frac2b}\arctan\left[\frac{\sqrt{2 b(b-2)}}{4\sqrt{\delta}}\right]\right)\\
&=\frac{\pi\omega}{2}-\sqrt{\frac2b}\frac{\pi\omega}{2}=(2-\sqrt{3+\xi})\frac{\pi\omega}{4}\nt.
\end{align*}
We therefore find that the total growth is given by
\eq{
\ln(\gamma^2)=\lim_{N\to\infty}2\,\mathcal I(N)\approx(2-\sqrt{3+\xi})\pi\omega.
}

To get mode functions in the regime of validity of the EFT, we want to evaluate
\eq{
\zeta\propto\exp\left[\pm(2-\sqrt{3+\xi})\frac{\pi\omega}4\mp F\left(\sqrt{1+\frac{16}{\omega^2(3+\xi)^2}\kappa^2}\right)\right],
}
in the limit $\kappa^2/\omega^2\ll1$. To do so, we need to compute $F(1+\delta)$ with $\delta=8\kappa^2/(3+\xi)^2\omega^2$ to the first non-trivial order in $\kappa$. Upon doing this, one finds
\begin{align*}
F(1+\delta)&\simeq F(1) +\frac{2\omega\sqrt{b-2}}b\sqrt{\delta}+\omega\frac{2\sqrt{\delta}}{\sqrt{b-2}}-\omega\frac{2\sqrt{\delta}}{\sqrt{b-2}}\\
&=F(1)+|c_s|\kappa
\end{align*}
where in the first line we used $\arctan(1/x)=\pi/2-\arctan(x)$ for $x>0$ and $\arctan(1/x)=-\pi/2-\arctan(x)$ for $x<0$, and in the second line we used  $|c_s|=\sqrt{(1-\xi)/(3+\xi)}$ and the definition of $b$. We therefore find the mode functions
\eq{
\zeta\propto\exp\left[\pm(2-\sqrt{3+\xi})\frac{\pi\omega}2\mp|c_s|\kappa\right].
}

\bibliographystyle{JHEP}
\bibliography{RTI2refs}

\end{document}